\documentclass[twocolumn,aps,superscriptaddress,prb]{revtex4}

\usepackage[T1]{fontenc}
\usepackage[scaled=0.92]{helvet}	

\usepackage[vvarbb]{newtxmath}
\usepackage{amsmath,amsfonts,bm}
\usepackage{graphicx}  
\usepackage{color}
\usepackage{extarrows}
\usepackage{enumitem}
\usepackage{empheq}
\usepackage{tensor}
\usepackage{esint}

\newcommand{\TT}{\mbox{$\stackrel{\leftrightarrow}{T}$}}
\newcommand{\Lpartial}{\mbox{$\stackrel{\leftarrow}{\partial}$}}
\newcommand{\Rpartial}{\mbox{$\stackrel{\rightarrow}{\partial}$}}

\renewcommand{\d}{{\rm d}}
\newcommand{\ud}{{\mathrm d}}

\newcommand{\w}{\omega}
\newcommand{\sbt}{SBET}
\newcommand{\wti}{\widetilde}
\newcommand{\ti}{\tilde}
\newcommand{\B}{\mbox{\tiny B}}

\newcommand{\tT}{\mbox{\tiny Tot}}

\newcommand{\tS}{{\mbox{\tiny S}}}
\newcommand{\tSS}{{\mbox{\tiny SS}}}
\newcommand{\T}{\mbox{\tiny T}}
\newcommand{\cla}{{\rm cl}}

\newcommand{\PB}{\mbox{\tiny PB}}
\newcommand{\dg}{\dagger}
\newcommand{\la}{\langle}
\newcommand{\ra}{\rangle}

\newcommand{\Sec}[1]{Sec.\,\ref{#1}}
\newcommand{\App}[1]{Appendix \ref{#1}}
\newcommand{\nl}{\nonumber \\}
\newcommand{\be}{\begin{equation}}
\newcommand{\ee}{\end{equation}}
\newcommand{\bsube}{\begin{subequations}}
\newcommand{\esube}{\end{subequations}}
\newcommand{\Eq}[1]{Eq.\,(\ref{#1})}
\newcommand{\Eqs}[1]{Eqs.\,(\ref{#1})}

\newcommand{\RN}[1]{%
  \textup{\uppercase\expandafter{\romannumeral#1}}%
}

\begin{document}
\title{Extended system-bath entanglement theorem for multiple bosonic or fermionic environments}
\author{Yu Su}
\affiliation{Hefei National Research Center for Physical Sciences at the Microscale,
University of Science and Technology of China, Hefei, Anhui 230026, China}
\author{Hao-Yang Qi}
\affiliation{Key Laboratory of Precision and Intelligent Chemistry, University of Science and Technology of China, Hefei 230026, China}
\author{Zi-Hao Chen}
\affiliation{Department of Chemistry, The University of Hong Kong, Pokfulam Road, Hong Kong, China}
\author{Yao Wang}     \email{wy2010@ustc.edu.cn}
\affiliation{Hefei National Research Center for Physical Sciences at the Microscale,
University of Science and Technology of China, Hefei, Anhui 230026, China}
\author{Rui-Xue Xu}   \email{rxxu@ustc.edu.cn}
\affiliation{Hefei National Research Center for Physical Sciences at the Microscale,
University of Science and Technology of China, Hefei, Anhui 230026, China}
\author{YiJing Yan}
\affiliation{Hefei National Research Center for Physical Sciences at the Microscale,
University of Science and Technology of China, Hefei, Anhui 230026, China}

\date{\today}

\begin{abstract}
The system-bath entanglement theorem (SBET) was established in terms of linear response functions [{\em J.~Chem.~Phys.~}{\bf 152}, 034102 (2020)] and generalized to correlation functions [arXiv:2312.13618 (2023)] in our previous works.
This theorem connects the entangled system--bath properties to the local system and bare bath ones.
In this work, firstly we extend the  \sbt\  
to field--dressed conditions with
multiple bosonic Gaussian environments at different temperatures.
Not only the system but also environments are considered to be of optical polarizability, as in reality.
With the aid of the extended SBET developed here, for the evaluation of the nonlinear spectroscopy such as the pump--probe, 
the entangled system--bath contributions can be obtained 
upon reduced system evolutions via certain quantum dissipative methods.
The extended SBET in the field-free condition and its counterpart in the classical limit
is also presented.
The SBET for fermionic environments is also elaborated within the transport scenarios for completeness.

\end{abstract}
\maketitle

\section{Introduction}

Open quantum systems play crucial roles in various modern research fields, where thermal baths may significantly affect the outcomes.\cite{Wei21} 
Developments of quantum dissipation theories (QDTs) are mainly based on Gaussian properties of baths, where the environments consist of non-interacting particles linearly coupled to the system.\cite{Cal83587,Gra88115}
The existing methods cover from perturbative master equation approaches\cite{Red651,Lin76393,yan002068} to the exact Feynman--Vernon influence functional path integral\cite{Fey63118} together with its differential equivalence, the hierarchical--equations--of--motion (HEOM) formalism.\cite{Tan906676,Yan04216,Xu05041103,Tan20020901}
Most QDTs focus on the reduced system dynamics.
However, the system-bath entanglements can be essential in such as Fano resonances,\cite{Fan611866,Mir102257} vibronic spectroscopies, and thermal transports.
To deal with the system–environment hybridized dynamics, the dissipaton--equation--of--motion (DEOM) approach, which is a second quantization generalization of the HEOM formalism, has been systematically developed in recent years.\cite{Yan14054105,Zha15024112,Xu151816,Wan22170901}

A system-bath entanglement theorem (\sbt) was established and explored in our previous work\cite{Du20034102,Su23Arxiv2312_13618} for arbitrary open quantum systems with the Gauss--Wick's coupling environment. 
We first proposed the SBET in terms of the linear response functions\cite{Du20034102} and later generalized it to correlation functions.\cite{Su23Arxiv2312_13618}
Both of them are established for steady states, not just thermal equilibrium states. 
The SBET connects the entangled system--bath properties in the total composite space to those of the local system and bare baths. 
It thus serves as a method to acquire the system--bath entangled information by   using only the system and bare--bath properties as the inputs. 
This theorem is established on the basis of the Langevin equation for the hybridizing bath modes.

In this work, we extend the \sbt\ to field--dressed conditions with multiple bosonic Gaussian environments at different temperatures.
In realistic scenarios, not only the system but also the environments can be excited in the presence of external fields.
We also elaborate the SBET with fermionic environments.
The extended SBET provides a universal framework to resolve the system--bath composite entanglement quantities contributed to such as quantum transport and pump--probe spectroscopy.\cite{Muk95, Yan898149, Shu013868}

The remainder of this paper is organized as follows. In \Sec{thsec2} and \ref{thsec3}, we present the bosonic and fermionic SBETs, respectively.
Theoretical background on the linear response theory is outlined in Appendix \ref{thappA}.
The classical limit of the bosonic field-free SBET is elaborated in Appendix \ref{thappB}.
Appendix \ref{thappC} introduces the related fermionic thermofield method, and  readers may consult Ref.\,\onlinecite{Su23Arxiv2312_13618} for its bosonic counterpart. 
Finally, we summarize the paper in \Sec{thsec4}.
Throughout this paper we set $ \hbar=1 $ and $ \beta_\alpha=1/(k_{B}T_{\alpha}) $ with  $ k_{B} $ being the Boltzmann constant and $ T_{\alpha} $ the temperature of the $\alpha$-reservoir.

\section{Extended system--bath entanglement theorem for bosonic bath}\label{thsec2}
\subsection{General description}
Let us start from the following total Hamiltonian
\begin{align}\label{HT}
    H_{\T}(t) &= H_{\tS} + \sum_{\alpha}h_{\alpha} - \sum_{v}\hat Q_v\big(\sum_{\alpha}\eta_{\alpha v}\hat F_{\alpha v}\big)
\nl &\quad
    -\Big[\sum_{v}\mu_v\hat Q_v\epsilon_v(t)+\sum_{\alpha,  v}\mu_{\alpha v}\hat F_{\alpha v}\epsilon_{\alpha v}(t)\Big],
\end{align}
where
\begin{align}\label{hB}
    \!\!h_{\alpha} = \frac{1}{2}\sum_j\omega_{\alpha j}(\hat p_{\alpha j}^2+\hat x_{\alpha j}^2)\quad \text{and}\quad \hat F_{\alpha v} = \sum_{j}c_{\alpha vj}\hat x_{\alpha j}.
\end{align}
Here, $H_{\tS}$ and $h_{\alpha}$ are the Hamiltonians of the system and the $\alpha$-environment, respectively. %
Each environment is composed of independent harmonic oscillators ($\{\hat x_{\alpha j}, \hat p_{\alpha j}\}$), which is equilibrated 
at the temperature $T_{\alpha}.$
The system modes $\{\hat Q_v\}$ are arbitrary Hermitian system operators, while $\{\hat F_{\alpha v}\}$ are the environmental hybridizing operators which are linear with respect to environmental particles' coordinates.
The last term in \Eq{HT} describes the interaction with the external classical  fields $\{\varepsilon_{v}(t), \varepsilon_{\alpha v}(t)\}$, where $\{\mu_v\}$ and $\{\mu_{\alpha v}\}$ are the dipolar strengths of the system and environments, respectively. 
It should be noticed that although all coupling terms are formally included in \Eq{HT},  we can adjust the elements of $\{\eta_{\alpha v}\}$, $\{\mu_v\}$ and $\{\mu_{\alpha v}\}$ in the system--environment, system--field and environment--field couplings in realistic applications.
For example, some of them can be zeros
for some $\alpha$ and $v$.

Consider the pump--probe spectroscopy.\cite{Muk95, Yan898149, Shu013868} The 
interested field--dressed correlation function is
\begin{align}\label{R12}
  R_{O_2O_1}(t_2,t_1)
  =i\,{\rm Tr}\Big[\hat O_2G_{\T}(t_2,t_1)\hat O_1G_{\T}(t_1,t_0)
 \rho_{\T}(t_0)G^\dg_{\T}(t_2,t_0)\Big],
\end{align}
where $G_{\T}(t_2,t_1)$ is the two-time-dependent propagator in the presence of external fields, satisfying
\be\begin{split}
\frac{\partial}{\partial t_2}G_{\T}(t_2,t_1)&=-iH_{\T}(t_2)G_{\T}(t_2,t_1),\\
\frac{\partial}{\partial t_1}G_{\T}(t_2,t_1)&=iG_{\T}(t_2,t_1)H_{\T}(t_1),
\end{split}
\ee
and $G^\dg_{\T}(t_2,t_1)G_{\T}(t_2,t_1)=G_{\T}(t_2,t_1)G^\dg_{\T}(t_2,t_1)=1$.
We have
\bsube\begin{align}
&\frac{\partial}{\partial t_2}R_{O_2O_1}(t_2,t_1)=
i\,{\rm Tr}\Big\{G^\dg_{\T}(t_2,t_0)i[H_{\T}(t_2),\hat O_2]
\nl &\qquad \times
 G_{\T}(t_2,t_1)
 \hat O_1G_{\T}(t_1,t_0)
 \rho_{\T}(t_0)\Big\},\\
&\frac{\partial}{\partial t_1}R_{O_2O_1}(t_2,t_1)=
i\,{\rm Tr}\Big\{G^\dg_{\T}(t_2,t_0)\hat O_2
 G_{\T}(t_2,t_1)
\nl &\qquad \times
 i[H_{\T}(t_1),\hat O_1]G_{\T}(t_1,t_0)
 \rho_{\T}(t_0)\Big\}.
\end{align}
\esube
In this work, the functions to be evaluated are for $\hat O_1$ or $\hat O_2$ $\in\{\{\hat Q_v\},\{\hat F_{\alpha v}\}\}$.

We can obtain
$\frac{\partial}{\partial t_2}R_{x_{\alpha j}O}(t_2,t_1)=\omega_{\alpha j}R_{p_{\alpha j}O}(t_2,t_1)$ and
\begin{align}\label{eq:6}
&\frac{\partial^2}{\partial t^2_2}R_{x_{\alpha j}O}(t_2,t_1)=-\omega^2_{\alpha j}R_{x_{\alpha j}O}(t_2,t_1)
\nl &\qquad 
+\sum_{v}\omega_{\alpha j}\eta_{\alpha v}c_{\alpha vj}R_{Q_vO}(t_2,t_1)
\nl &\qquad +i\sum_{v}\omega_{\alpha j}\mu_{\alpha v}c_{\alpha vj}\epsilon_{\alpha v}(t_2)\overline{O(t_1)},
\end{align}
where we denote $\overline{O(t)}\equiv {\rm Tr}[\hat O\rho_{\T}(t)]$ with $\rho_{\T}(t) = G_{\T}(t,t_0)\rho_{\T}(t_0)G^\dg_{\T}(t,t_0)$.
The solution to \Eq{eq:6} can be formally obtained as
\begin{align}
&R_{x_{\alpha j}O}(t_2,t_1)=R_{x_{\alpha j}O}(t_1,t_1)\cos[\omega_{\alpha j}(t_2-t_1)]
\nl &\qquad +R_{p_{\alpha j}O}(t_1,t_1)\sin[\omega_{\alpha j}(t_2-t_1)]
\nl &\qquad +\sum_{v}\int_{t_1}^{t_2}\!\d\tau\,c_{\alpha vj}\sin[\omega_{\alpha j}(t_2-\tau)]
\nl &\qquad\quad \times[\eta_{\alpha v}R_{Q_vO}(\tau,t_1)
+i\mu_{\alpha v}\epsilon_{\alpha v}(\tau)\overline{O(t_1)}].
\end{align}
Introduce 
\begin{align}\label{tiF}
\ti F^{\B}_{\alpha v}(t)
=\sum_{j}c_{\alpha vj}[\hat x_{\alpha j}\cos(\omega_{\alpha j}t)
  +\hat p_{\alpha j}\sin(\omega_{\alpha j}t)],
  \end{align}
  and
\be 
    \phi_{\alpha vv'}(t) = \sum_{j}c_{\alpha vj}c_{\alpha v'j}\sin(\omega_{\alpha j}t)=\phi_{\alpha v'v}(t),
\ee
together with
$\overline{F^{\B}_{\alpha v}(t')O(t)}\equiv {\rm Tr}[\ti F^{\B}_{\alpha v}(t')\hat O
    \rho_{\T}(t)]$.
We further obtain 
\begin{align}\label{eq12}
&R_{F_{\alpha v}O}(t_2,t_1)=i\overline{F^{\B}_{\alpha v}(t_2-t_1)O(t_1)}
\nl &\quad +\sum_{v'}\int_{t_1}^{t_2}\!\d\tau\,\phi_{\alpha vv'}(t_2-\tau)
 [\eta_{\alpha v'}R_{Q_{v'}O}(\tau,t_1)
\nl &\qquad\qquad
  +i\mu_{\alpha v'}\epsilon_{\alpha v'}(\tau)\overline{O(t_1)}].
\end{align}
Similarly, there are
$\frac{\partial}{\partial t_1}R_{Ox_{\alpha j}}(t_2,t_1)=\omega_{\alpha j}R_{Op_{\alpha j}}(t_2,t_1)$ and
\begin{align}
&\frac{\partial^2}{\partial t^2_1}R_{Ox_{\alpha j}}(t_2,t_1)=-\omega^2_{\alpha j}R_{Ox_{\alpha j}}(t_2,t_1)
\nl &\qquad +\sum_{v}\omega_{\alpha j}\eta_{\alpha v}c_{\alpha vj}R_{OQ_v}(t_2,t_1)
\nl &\qquad +i\sum_{v}\omega_{\alpha j}\mu_{\alpha v}c_{\alpha vj}\epsilon_{\alpha v}(t_1)\overline{O(t_2)},
\end{align}
with the solution
\begin{align}
&R_{Ox_{\alpha j}}(t_2,t_1)=R_{Ox_{\alpha j}}(t_2,t_2)\cos[\omega_{\alpha j}(t_2-t_1)]
\nl &\qquad +R_{Op_{\alpha j}}(t_2,t_2)\sin[\omega_{\alpha j}(t_2-t_1)]
\nl &\qquad +\sum_{v}\int_{t_1}^{t_2}\!\d\tau\,c_{\alpha vj}\sin[\omega_{\alpha j}(\tau-t_1)]
\nl &\qquad\quad \times[\eta_{\alpha v}R_{OQ_v}(t_2,\tau)
+i\mu_{\alpha v}\epsilon_{\alpha v}(\tau)\overline{O(t_2)}],
\end{align}
leading to
\begin{align}\label{eq15}
&R_{OF_{\alpha v}}(t_2,t_1)=i\overline{O(t_2)F^{\B}_{\alpha v}(t_2-t_1)}
\nl &\quad +\sum_{v'}\int_{t_1}^{t_2}\!\d\tau\,\phi_{\alpha vv'}(\tau-t_1)
 [\eta_{\alpha v'}R_{OQ_{v'}}(t_2,\tau)
\nl &\qquad\qquad
  +i\mu_{\alpha v'}\epsilon_{\alpha v'}(\tau)\overline{O(t_2)}],
\end{align}
where
$\overline{O(t)F^{\B}_{\alpha v}(t')}\equiv {\rm Tr}[\hat O\ti F^{\B}_{\alpha v}(t')\rho_{\T}(t)]$.
Equations (\ref{eq12}) and (\ref{eq15}) are the precursor of SBET,  where $\overline{F^{\B}_{\alpha v}(t)O(t')}$ and $\overline{O(t')F^{\B}_{\alpha v}(t)}$ $(\hat O \in\{\{\hat Q_v\},\{\hat F_{\alpha v}\}\})$
together with 
$\overline{O(t)}$ $(\hat O \in\{\hat F_{\alpha v}\})$
shall be further determined.
The $R_{Q_{v}Q_{v'}}(t,t')$ and $\overline{Q_v(t)}$ are related to reduced system dynamics and can be obtained via a certain QDT method.

\subsection{Derivations}
\subsubsection{Prelude on thermofield decomposition}
This subsection deals with $\overline{F^{\B}_{\alpha v}(t)Q_{v'}(t')}$,
$\overline{Q_{v'}(t')F^{\B}_{\alpha v}(t)}$,
$\overline{F^{\B}_{\alpha v}(t)F^{\B}_{\alpha' v'}(t')}$
 and
$\overline{F_{\alpha v}(t)}$.
To do this, we consult the thermofield approach.\cite{Ume95} 
In this approach, we need to define the operators in the Heisenberg picture as 
\begin{align}\label{Gnorm}
    \hat O(t) \equiv G_{\T}^\dagger(t,t_0)\hat O G_{\T}(t,t_0),
\end{align}
and then \Eq{R12} shall be expressed in the Heisenberg picture as
\begin{align}\label{eq17}
    R_{O_2O_1}(t_2,t_1) \equiv  
    i\,{\rm Tr}\big[ \hat O_2(t_2)\hat O_1(t_1)\rho_{\T}(t_0) \big].
\end{align}
The Heisenberg equation reads
\begin{align}
    \frac{\ud}{\ud t}\hat O(t) = i G_{\T}^\dagger(t,t_0) [H_{\T}(t),\hat O] G_{\T}(t,t_0). 
\end{align}
Then, we have 
\begin{subequations}
    \begin{align}
        \frac{\ud}{\ud t}\hat x_{\alpha j}(t) &= \omega_{\alpha j}\hat p_{\alpha j}(t),\\
        \frac{\ud}{\ud t}\hat p_{\alpha j}(t) &= -\omega_{\alpha j}\hat x_{\alpha j}(t) + \sum_v c_{\alpha vj}\Big[ \eta_{\alpha v}\hat Q_v(t)+\mu_{\alpha v}\varepsilon_{\alpha v}(t) \Big].
    \end{align}
\end{subequations}
Given initial conditions $\hat x_{\alpha j}(t_0) = \hat x_{\alpha j}$ and $\hat p_{\alpha j}(t_0) = \hat p_{\alpha j}$, one can readily obtain 
\begin{align}
    \quad\hat x_{\alpha j}(t) &= \hat x_{\alpha j}^{\B}(t) + \sum_v c_{\alpha vj}\int_{t_0}^t\!\!\ud\tau\sin[\omega_{\alpha j}(t-\tau)] \nl  &\qquad\qquad\quad\times\Big[\eta_{\alpha v}\hat Q_v(\tau) + \mu_{\alpha v}\varepsilon_{\alpha v}(\tau)\Big].
\end{align}
Here we introduce the operators in the bare--bath interaction picture, $\hat O^{\B}(t) \equiv e^{ih_{\B}(t-t_0)}\hat Oe^{-ih_{\B}(t-t_0)}$ with $h_{\B}=\sum_{\alpha} h_{\alpha}$. 
According to \Eq{hB}, we obtain
\begin{align}\label{eq21}
    \hat F_{\alpha v}(t) = \hat F_{\alpha v}^{\B}(t) &+ \sum_{v'}\int_{t_0}^t\!\ud\tau\,\phi_{\alpha vv'}(t-\tau)\nl 
    &\qquad\times\Big[\eta_{\alpha v'} \hat Q_{v'}(\tau)+\mu_{\alpha v'}\varepsilon_{\alpha v'}(\tau) \Big].
\end{align}
Note that $\hat F^{\B}_{\alpha v}(t)=\ti F^{\B}_{\alpha v}(t-t_0)$ [cf.\,\Eq{tiF}].

The thermofield decomposition for the initial factorized state $\rho_{\T}(t_0)=\rho_{\tS}(t_0)\otimes \prod_{\alpha} e^{-\beta_{\alpha}h_{\alpha}}/{\rm tr}_{\alpha}(e^{-\beta_{\alpha}h_{\alpha}})$ reads
\begin{align}\label{eq22}
    \hat F_{\alpha v}(t) = \hat F^+_{\alpha v}(t) + \hat F^-_{\alpha v}(t),
\end{align}
where $\hat F^+_{\alpha v}(t) = [\hat F^-_{\alpha v}(t)]^\dagger$.
Concerning the steady state before the action of the external field,  $t_0$ is effectively set to be $-\infty$. Equation (\ref{eq17}) is recast to
\begin{align}\label{correff2}
   R_{O_2O_1}(t_2,t_1)
    =i\,{\wti{\rm Tr}}\big[&
   \wti O_2(t_2)\wti O_1(t_1)\rho_{\tS}(t_0)|\xi\ra\la\xi|\big],
\end{align}
with 
\begin{align}\label{Gintherm}
    \wti O(t)\equiv \wti G_{\T}^\dagger(t,t_0)\hat O \wti G_{\T}(t,t_0).
\end{align}
Here $\wti G_{\T}(t,t_0)$ is the propagator according to the Hamiltonian $\wti H_{\T}(t)=H_{\T}(t)+h'_{\B}$ with $h'_{\B}$ being the auxiliary bath Hamiltonian.
In \Eq{correff2}, $\wti{\rm Tr}(\cdots)={\rm Tr}[{\rm tr}_{\B}'(\cdots)]$ is the trace over the entire space of $\wti H_{\T}(t)$. $|\xi\ra$ is denoted as
the vacuum state of the effective bath $\ti h_{\B}=h_{\B}+h'_{\B}$. 
In \Eq{eq22}
\begin{align}\label{Fintherm}
\hat F^{\pm}_{\alpha v}(t) = \hat F^{\pm;\B}_{\alpha v}(t) &\mp i\sum_{v'} \int_{t_0}^t\!\ud\tau\, c^{\pm}_{\alpha vv'}(t-\tau)\nl 
    &\qquad\times\Big[\eta_{\alpha v'}\hat Q_{v'}(\tau)+\mu_{\alpha v'}\varepsilon_{\alpha v'}(\tau)\Big],
\end{align}
where $c_{\alpha vv'}^+(t) = [c^-_{\alpha vv'}(t)]^*$,
\begin{align}\label{FDT}
    c^-_{\alpha vv'}(t) = \frac{1}{2\pi i}\int_{-\infty}^\infty\!\!\ud\omega\frac{e^{-i\omega t}}{1-e^{-\beta\omega}}\int_{-\infty}^\infty\!\!\ud\tau\,e^{i\omega\tau}\phi_{\alpha vv'}(\tau),
\end{align}
and
\begin{align}\label{Fxi}
    \hat F^{-;\B}_{\alpha v}(t)|\xi\ra\la\xi| = |\xi\ra\la\xi|\hat F^{+;\B}_{\alpha v}(t) = 0.
\end{align}
Equation (\ref{FDT}) is the fluctuation--dissipation theorem.\cite{Wei21,Yan05187}
Note that in \Eq{Fintherm}, $\hat F^{\pm}_{\alpha v}(t)$ defined as \Eq{Gnorm} equals to that defined as \Eq{Gintherm} and $\hat F^{\pm;\B}_{\alpha v}(t)$ similar.
More details can be found in Ref.\,\onlinecite{Su23Arxiv2312_13618}.

\subsubsection{$R_{F_{\alpha v}Q_{v'}}(t_2,t_1)$}

From \Eq{correff2},
\begin{align}\label{28}
    &\quad R_{F_{\alpha v}Q_{v'}}(t_2,t_1)\nl 
    &= 2i\,{\rm Re}\wti{\rm Tr}\Big[\hat F_{\alpha v}^+(t_2)\hat Q_{v'}(t_1)\rho_{\tS}(t_0)|\xi\ra\la \xi|\Big] \nl
    &\quad + i\,\wti{\rm Tr}\Big\{\big[\hat F^-_{\alpha v}(t_2),\hat Q_{v'}(t_1)\big]\rho_{\tS}(t_0)|\xi\ra\la \xi|\Big\} \nl 
    &\equiv i\,[{\rm (I)+(II)}].
\end{align}
Substituting \Eq{Fintherm} into \Eq{28} together with \Eq{Fxi}, we obtain
\begin{align}\label{29}
    {\rm (I)} = -2{\rm Re} \sum_{v''}&\int_{t_0}^{t_2}\!\ud\tau\,c^+_{\alpha vv''}(t_2-\tau)\Big[\eta_{\alpha v''}R_{Q_{v''}Q_{v'}}(\tau,t_1) \nl 
    &\quad +i \mu_{\alpha v''}\varepsilon_{\alpha v''}(\tau)\overline{{Q}_{v'}(t_1)}\Big].
\end{align}
To evaluate $\rm (II)$, we notice that \Eq{Fintherm} can be recast as 
\begin{align}\label{Fintherm1}
\hat F^{\pm}_{\alpha v}(t_2) = \hat X^{\pm;\B}_{\alpha v}(t_2) &\mp i\sum_{v'} \int_{t_1}^{t_2}\!\ud\tau\, c^{\pm}_{\alpha vv'}(t_2-\tau)\nl 
    &\quad\times\Big[\eta_{\alpha v'}\hat Q_{v'}(\tau)+\mu_{\alpha v'}\varepsilon_{\alpha v'}(\tau)\Big],
\end{align}
where $\hat X^{\pm;\B}_{\alpha v}(t_2) \equiv \wti G_{\T}^\dagger(t_1,t_0)e^{i\ti h_{\B}(t_2-t_1)}\hat F^\pm_{\alpha v}e^{-i\ti h_{\B}(t_2-t_1)}\wti G_{\T}(t_1,t_0)$ commutes with $\hat Q_{v'}(t_1)$. Thus
\begin{align}\label{31}
    {\rm (II)} =  \sum_{v''}\int_{t_1}^{t_2}\!\ud\tau\, c^-_{\alpha vv''}(t_2-\tau)&\eta_{\alpha v''}\Big[ R_{Q_{v''}Q_{v'}}(\tau, t_1) \nl 
    &\quad -R_{Q_{v'}Q_{v''}}(t_1, \tau) \Big]. 
\end{align}
We thus finish \Eq{28} by combining \Eqs{29} and (\ref{31}). Compared with \Eq{eq12}, we obtain 
\begin{align}\label{FbQ}
    &\quad\overline{F_{\alpha v}^{\B}(t_2-t_1)Q_{v'}(t_1)} = -2{\rm Re}\Bigg\{ \sum_{v''}\int_{t_0}^{t_1}\!\ud\tau\,c^+_{\alpha vv''}(t_2-\tau)\nl 
    &\quad\times\Big[\eta_{\alpha v''}R_{Q_{v''}Q_{v'}}(\tau,t_1) +i \mu_{\alpha v''}\varepsilon_{\alpha v''}(\tau)\overline{{Q}_{v'}(t_1)}\Big]\Bigg\}.
\end{align}
We see that the underlying system--bath entanglement  in the l.h.s of \Eq{FbQ} is accounted for by the memory integration convolved between the bare--bath correlation and the system dynamics. 

\subsubsection{$R_{Q_{v'}F_{\alpha v}}(t_1,t_2)$}

From \Eq{correff2},
\begin{align}\label{33}
    &\quad R_{Q_{v'}F_{\alpha v}}(t_1,t_2)\nl 
    &= 2i\,{\rm Re}\wti{\rm Tr}\Big[\hat Q_{v'}(t_1)\hat F^-_{\alpha v}(t_2)\rho_{\tS}(t_0)|\xi\ra\la \xi|\Big] \nl
    &\quad + i\,\wti{\rm Tr}\Big\{\big[\hat Q_{v'}(t_1),\hat F^+_{\alpha v}(t_2)\big]\rho_{\tS}(t_0)|\xi\ra\la \xi|\Big\} \nl 
    &\equiv i\,[{\rm (I')+(II')}].
\end{align}
Substituting \Eq{Fintherm} into \Eq{33} together with \Eq{Fxi}, we obtain
\begin{align}\label{34}
    {\rm (I')} = 2{\rm Re} \sum_{v''}&\int_{t_0}^{t_2}\!\ud\tau\,c^-_{\alpha vv''}(t_2-\tau)\Big[\eta_{\alpha v''}R_{Q_{v'}Q_{v''}}(t_1,\tau) \nl 
    &\quad +i \mu_{\alpha v''}\varepsilon_{\alpha v''}(\tau)\overline{{Q}_{v'}(t_1)}\Big].
\end{align}
With \Eq{Fintherm1},
\begin{align}\label{35}
    \!\!{\rm (II')} =  -\sum_{v''}\int_{t_1}^{t_2}\!\ud\tau\, c^+_{\alpha vv''}(t_2-\tau)&\eta_{\alpha v''}\Big[ R_{Q_{v'}Q_{v''}}(t_1, \tau) \nl 
    &-R_{Q_{v''}Q_{v'}}(\tau, t_1)\Big]. 
\end{align}
We thus finish \Eq{33} by combining \Eqs{34} and (\ref{35}). Compared with \Eq{eq15}, we obtain 
\begin{align}\label{QFb}
    &\quad\overline{Q_{v'}(t_1)F_{\alpha v}^{\B}(t_1-t_2)} = 2{\rm Re} \Bigg\{\sum_{v''}\int_{t_0}^{t_1}\!\ud\tau\,c^-_{\alpha vv''}(t_2-\tau)\nl 
    &\quad\times\Big[\eta_{\alpha v''}R_{Q_{v'}Q_{v''}}(t_1,\tau) +i \mu_{\alpha v''}\varepsilon_{\alpha v''}(\tau)\overline{{Q}_{v'}(t_1)}\Big]\Bigg\}.
\end{align}
Compared  to \Eq{FbQ}, we find $\overline{F^{\B}_{\alpha v}(t)Q_{v'}(t')}=\overline{Q_{v'}(t')F^{\B}_{\alpha v}(-t)}$. By the definitions before \Eq{eq12} and after \Eq{eq15} together with \Eq{tiF}, $\overline{Q_{v'}(t')F^{\B}_{\alpha v}(-t)}=\overline{F^{\B}_{\alpha v}(-t)Q_{v'}(t')}$.
Therefore $\overline{F^{\B}_{\alpha v}(t)Q_{v'}(t')}=\overline{F^{\B}_{\alpha v}(-t)Q_{v'}(t')}$. This result arises from the nature of $\ti F^{\B}_{\alpha v}(t)$ as a random force.

\subsubsection{$R_{F_{\alpha v}F_{\alpha'v'}}(t_2,t_1)$}

From \Eq{correff2},
\begin{align}\label{eq37}
    &\quad R_{F_{\alpha v}F_{\alpha'v'}}(t_2,t_1)\nl 
    &= 2i\,{\rm Re}\wti{\rm Tr}\Big[ \hat F^+_{\alpha v}(t_2)\hat F_{\alpha'v'}(t_1)\rho_{\tS}(t_0)|\xi\ra\la\xi| \Big] \nl
    &\quad + i\,\wti{\rm Tr}\Big\{\big[ \hat F^-_{\alpha v}(t_2),\hat F_{\alpha'v'}(t_1) \big]\rho_{\tS}(t_0)|\xi\ra\la \xi|\Big\} \nl 
    &\equiv i\,[{\rm (I'')+(II'')}].
\end{align}
Substituting \Eq{Fintherm} into \Eq{33} together with \Eq{Fxi}, we obtain
\begin{align}\label{eq38}
    {\rm (I'')} = -2{\rm Re} \sum_{v''}&\int_{t_0}^{t_2}\!\ud\tau\,c^+_{\alpha vv''}(t_2-\tau)\Big[\eta_{\alpha v''}R_{Q_{v''}F_{\alpha' v'}}(\tau,t_1) \nl 
    &\quad +i \mu_{\alpha v''}\varepsilon_{\alpha v''}(\tau)\overline{{F}_{\alpha'v'}(t_1)}\Big].
\end{align}
With \Eq{Fintherm1},
\begin{align}\label{eq39}
    \!\!{\rm (II'')} &= \delta_{\alpha\alpha'}c_{\alpha vv'}(t_2-t_1) +\sum_{v''}\int_{t_1}^{t_2}\!\ud\tau\, c^-_{\alpha vv''}(t_2-\tau)\eta_{\alpha v''}\nl 
    &\qquad\quad\times\Big[ R_{Q_{v''}F_{\alpha'v'}}(\tau,t_1) -R_{F_{\alpha'v'}Q_{v''}}(t_1, \tau)\Big]. 
\end{align}
The first term is obtained from the thermofield decomposition as detailed in Ref.\,\onlinecite{Su23Arxiv2312_13618}.
We thus finish \Eq{eq37} by combining \Eqs{eq38} and (\ref{eq39}). Compared with \Eq{eq12}, we obtain 
\begin{align}\label{FbFb}
    &\quad\overline{F^{\B}_{\alpha v}(t_2-t_1)F_{\alpha'v'}(t_1)}\nl 
    &= \delta_{\alpha\alpha'}c_{\alpha vv'}(t_2-t_1) - 2{\rm Re}\Bigg\{ \sum_{v''}\int_{t_0}^{t_1}\!\ud\tau\,c^+_{\alpha vv''}(t_2-\tau)\nl 
    &\times\Big[\eta_{\alpha v''}R_{Q_{v''}F_{\alpha'v'}}(\tau,t_1) +i \mu_{\alpha v''}\varepsilon_{\alpha v''}(\tau)\overline{{F}_{\alpha'v'}(t_1)}\Big]\Bigg\}.
\end{align}
In \Eq{FbFb}, 
\be 
\overline{{F}_{\alpha v}(t)}=\sum_{v'}\int_{t_0}^t\!\ud\tau\,\phi_{\alpha vv'}(t-\tau)
\Big[\eta_{\alpha v'} \overline{ Q_{v'}(\tau)}+\mu_{\alpha v'}\varepsilon_{\alpha v'}(\tau) \Big]
\ee
according to \Eq{eq21}.
This indicates the actions of the system mode  and the external field onto the bath.
Due to the Gaussian nature of the bath, nonlinear responses vanish. 

Up to now, we have  completed the extended SBET of multi--bosonic baths polarized by external fields. 
As long as  $R_{Q_{v}Q_{v'}}(t,t')$ and $\overline{Q_v(t)}$  related to reduced system dynamics are obtained via a certain QDT method, the entangled system--bath and bath--bath properties can then be evaluated.
It is worth reemphasizing that although $\sum_{\alpha,v}\eta_{\alpha v}\hat Q_v\hat F_{\alpha v}$ are formally included in \Eq{HT},  we can set some elements in $\{\eta_{\alpha v}\}$, $\{\mu_v\}$ and $\{\mu_{\alpha v}\}$ in the system--environment, system--field and environment--field interaction terms to be zeros when particular couplings are absent.
In such cases,
the extended SBET is still valid.

\section{System--bath entanglement theorem for fermionic bath}\label{thsec3}

\subsection{Prelude}
The fermionic transport is anticipated when the local impurity electronic system is coupled to multiple electrodes with different temperatures and/or chemical potentials. The total composite Hamiltonian reads 
\begin{align}\label{fermion}
H_{\T} = H_{\tS} (\{\hat\psi_u^{\dg}, \hat\psi_u\})\!+ \!\sum_{\alpha}h_{\alpha}^{\rm eff} \!+ \sum_{\alpha u}(\hat\psi_u^\dagger\hat F_{\alpha u}\!+\!\hat F_{\alpha u}^\dagger\hat\psi_u),
\end{align}
with the effective bath Hamiltonian
\begin{subequations}\label{fermion1}
    \begin{align}
    h_{\alpha}^{\rm eff} = \sum_j(\epsilon_{\alpha j}-\mu_{\alpha})\hat a_{\alpha j}^\dagger\hat a_{\alpha j} \equiv h_{\alpha}-\mu_{\alpha}\hat N_{\alpha}
    \end{align}
    and 
    \begin{align}\label{Fau}
        \hat F_{\alpha u}=\sum_j t_{\alpha uj}^{\ast}\hat a_{\alpha j}.
    \end{align}
\end{subequations}
Here, the parameter $t_{\alpha uj}$ denotes the coupling strength between the central impurity (system) and  the reservoirs (baths).
In \Eq{fermion}, the local system $ H_{\tS} (\{\hat\psi_u^{\dg}, \hat\psi_u\})$ is arbitrary, containing often open--shell electrons with
strong Coulomb interactions. 
Similar as the derivation to the bosonic SBET, we adopt the thermofield decomposition as,
\begin{align}\label{48}
 \hat F_{\alpha u}= \hat F^{(1)}_{\alpha u}+\hat F^{(2)\dg}_{\alpha u}
\end{align}
with
\bsube
\be
\hat F^{(1)}_{\alpha u}=\sum_j t^{*}_{\alpha u j}\sqrt{1-\bar{n}_{\alpha j}}\hat d'_{\alpha j},
\ee
and
\be
\hat F^{(2)}_{\alpha u}=\sum_j t_{\alpha u j}\sqrt{\bar{n}_{\alpha j}}\hat d_{\alpha j}.
\ee
\esube
Details of the thermofield decomposition of non-interacting fermions are given in \App{thappC}.
We assume the initial state given by the factorization ansatz, namely 
\begin{align}
    \rho_{\T}(t_0) = \rho_{\tS}(t_0)\otimes\rho_{\B}^{\rm eff}.
\end{align}
Here $\rho_{\B}^{\rm eff}\equiv \prod_\alpha\otimes e^{-\beta h_\alpha^{\rm eff}}/{\rm tr}_ae^{-\beta h_\alpha^{\rm eff}}$ can be 
purified by a vacuum state denoted as $|\xi\ra$ via adding auxiliary baths; see \App{thappC}.

In the Heisenberg picture, the operator evolves as 
\be\hat O(t)\equiv e^{iH_{\T}(t-t_0)}\hat Oe^{-iH_{\T}(t-t_0)}.
\ee
We can obtain 
\begin{align}
    \hat{a}_{\alpha j}(t) =&\ \hat a_{\alpha j}^{\B}(t)e^{i\mu_{\alpha}(t-t_0)}\nl 
    &-i\sum_u\int_{t_0}^t\!\ud\tau\, t_{\alpha uj}e^{-i(\epsilon_{\alpha j}-\mu_\alpha)(t-\tau)}\hat\psi_{u}(\tau),
\end{align}
where $\hat a_{\alpha j}^{\B}(t)\equiv e^{ih_{\B}(t-t_0)}\hat a_{\alpha j}e^{-ih_{\B}(t-t_0)}$ with $h_{\B}\equiv\sum_\alpha h_\alpha$. 
According to \Eq{Fau}, 
we have 
\begin{align}\label{45}
    \hat F_{\alpha u}(t) =&\ \hat F_{\alpha u}^{\B}(t)e^{i\mu_{\alpha}(t-t_0)}\nl 
    &-i\sum_v\int_{t_0}^t\!\!\ud\tau\, g_{\alpha uv}(t-\tau)e^{i\mu_{\alpha}(t-\tau)}\hat\psi_v(\tau),
\end{align}
where $\hat F_{\alpha u}^{\B}(t)\equiv e^{ih_{\B}(t-t_0)}\hat F_{\alpha u}e^{-ih_{\B}(t-t_0)}$ and
\begin{align}
    g_{\alpha uv}(t)\equiv\la\{\hat F_{\alpha u}^{\B}(t),\hat F_{\alpha v}^{\B\dagger}(0)\}\ra_{\B},
\end{align}
with $\la\cdots\ra_{\B}\equiv {\rm tr}_{\B}[\cdots\rho_{\B}^{\rm eq}]$ and $\rho_{\B}^{\rm eq}\equiv e^{-\beta h_{\B}}/{\rm tr}_{\B}e^{-\beta h_{\B}}$.
We have
\begin{subequations}
\begin{align}
\hat d_{\alpha j}(t) =&\ \hat d_{\alpha j}^{\B}(t)e^{-i\mu_{\alpha}(t-t_0)} \nl 
&+i\sum_{u}\int_{t_0}^t\!\!\ud\tau\,t_{\alpha uj}^*\sqrt{\bar n_{\alpha j}}e^{i(\epsilon_{\alpha j}-\mu_{\alpha})(t-\tau)}\hat \psi_u^\dagger(\tau),
\\
\hat d'_{\alpha j}(t) =&\ \hat d_{\alpha j}^{\prime\B}(t)e^{i\mu_{\alpha}(t-t_0)} \nl 
&-i\sum_{u}\int_{t_0}^t\!\!\ud\tau\,t_{\alpha uj}\sqrt{1-\bar n_{\alpha j}}e^{-i(\epsilon_{\alpha j}-\mu_{\alpha})(t-\tau)}\hat \psi_u(\tau),
    \end{align}
\end{subequations}
and thus
\begin{subequations}\label{Ffermi}
\begin{align}\label{51a}
 \hat F_{\alpha u}^{(1)}(t) =&\ \hat F_{\alpha u}^{(1)\B}(t)e^{i\mu_{\alpha}(t-t_0)}\nl 
     &-i\sum_{v}\int_{t_0}^t\!\!\ud\tau\, c_{\alpha uv}^{(-)}(t-\tau)e^{i\mu_{\alpha}(t-\tau)}\hat \psi_v(\tau),\\ 
     \label{51b}
        \hat F_{\alpha u}^{(2)}(t) =&\ \hat F_{\alpha u}^{(2)\B}(t)e^{-i\mu_{\alpha}(t-t_0)}\nl 
        &+i\sum_{v}\int_{t_0}^t\!\!\ud\tau\, c_{\alpha uv}^{(+)}(t-\tau)e^{-i\mu_{\alpha}(t-\tau)}\hat \psi_v^\dagger(\tau).
    \end{align}
\end{subequations}
Here we define 
\begin{subequations}
\begin{align}
&c^{(-)}_{\alpha uv}(t)\equiv \la\hat F_{\alpha u}^{\B}(t)\hat F_{\alpha v}^{\B\dagger}(0)\ra_{\B}, \\
&c^{(+)}_{\alpha uv}(t)\equiv \la\hat F_{\alpha u}^{\B\dagger}(t)\hat F_{\alpha v}^{\B}(0)\ra_{\B},
    \end{align}
\end{subequations}
satisfying 
\begin{align}
    g_{\alpha uv}(t) = c^{(-)}_{\alpha uv}(t) + c^{(+)*}_{\alpha uv}(t).
\end{align}
Apparently, \Eqs{51a} and (\ref{51b}) together with \Eq{48} reproduce  \Eq{45}.
It is worth noticing that 
\begin{align}\label{fermivacuum}
    \hat F_{\alpha u}^{(1)\B}(t)|\xi\ra\la\xi| = \hat F_{\alpha u}^{(2)\B}(t)|\xi\ra\la\xi|=0, 
\end{align}
where $|\xi\ra\la\xi|$ is the effective vacuum [cf.\,\Eq{tbf1}].

\subsection{Derivations}
In this subsection, we construct the SBET for electronic impurity systems. We define the ensemble average with  respect to the  steady state, 
\begin{align}
    \rho^{\rm st} \equiv \lim_{t_0\to-\infty}e^{iH_{\T}t_0}\rho_{\T}(t_0)e^{-iH_{\T}t_0}.
\end{align}
Therefore for arbitrary correlation functions we have 
\begin{align}
    \la\hat A(t)\hat B(0)\ra 
    &\equiv {\rm Tr}[e^{iH_{\T}t}
       \hat Ae^{-iH_{\T}t}\hat B\rho^{\rm st}]     \nl 
    &= \lim_{t_0\to-\infty}{\rm Tr}
      [\hat A(t)\hat B(0)\rho_{\T}(t_0)]
\nl 
    &= \lim_{t_0\to-\infty}\wti{\rm Tr} 
      [\hat A(t)\hat B(0)\rho_{\tS}(t_0)|\xi\ra\la \xi|].
\end{align}
We denote the system correlations
\begin{subequations}
    \begin{align}\label{57a}
    &C_{uv}^{\tSS (-)}(t)\equiv\la\hat \psi_u(t)\hat \psi_v^\dagger(0)\ra,\\ \label{57b}
    &C_{uv}^{\tSS (+)}(t)\equiv\la\hat \psi_u^\dagger(t)\hat \psi_v(0)\ra,
    \end{align}
\end{subequations}
and the system--bath correlations 
\begin{subequations}
    \begin{align}
        \label{58a}&C_{uv}^{\alpha\tS(-)}(t)\equiv\la\hat F_{\alpha u}(t)\hat \psi_v^\dagger(0)\ra, \\ 
        \label{58b}&C_{uv}^{\alpha\tS(+)}(t)\equiv\la\hat F_{\alpha u}^\dagger(t)\hat \psi_v(0)\ra, \\ 
        \label{58c}&C_{uv}^{\tS\alpha(-)}(t)\equiv\la\hat \psi_u(t)\hat F_{\alpha v}^\dagger(0)\ra, \\
        \label{58d}&C_{uv}^{\tS\alpha(+)}(t)\equiv\la\hat \psi_u^\dagger(t)\hat F_{\alpha v}(0)\ra,
    \end{align}
\end{subequations}
with 
\begin{align}
    C_{uv}^{\alpha\tS(\pm)}(t) = [C_{vu}^{\tS\alpha(\pm)}(-t)]^*,
\end{align}
and the bath--bath correlations
\begin{subequations}
    \begin{align}
    \label{60a}&C_{uv}^{\alpha\alpha'(-)}(t)\equiv\la\hat F_{\alpha u}(t)\hat F_{\alpha'v}^\dagger(0)\ra,\\
    \label{60b}
    &C_{uv}^{\alpha\alpha'(+)}(t)\equiv\la\hat F_{\alpha u}^\dagger(t)\hat F_{\alpha' v}(0)\ra.
    \end{align}
\end{subequations}

For $C_{uv}^{\alpha\tS(-)}(t)$, we recast it as 
\begin{align}\label{61}
    C_{uv}^{\alpha\tS(-)}(t) &=\la \hat F_{\alpha u}^{(2)\dagger}(t)\hat \psi_v^\dagger(0)\ra-\la \hat \psi_v^\dagger(0)\hat F_{\alpha u}^{(1)}(t)\ra \nl 
    &\quad +\la\{ \hat F_{\alpha u}^{(1)}(t),\hat \psi_v^\dagger(0)\}\ra.
\end{align}
Substituting \Eq{Ffermi} into (\ref{61}) together with \Eq{fermivacuum}, we obtain 
\begin{align}
    &\quad C_{uv}^{\alpha\tS(-)}(t)\nl 
    &= -i\sum_{u'}\int_{-\infty}^t\!\!\ud\tau\,c_{\alpha uu'}^{(+)*}(t-\tau)e^{i\mu_{\alpha}(t-\tau)}C_{u'v}^{\tSS(-)}(\tau)\nl 
    &\quad +i\sum_{u'}\int_{-\infty}^t\!\!\ud\tau\,c_{\alpha uu'}^{(-)}(t-\tau)e^{i\mu_\alpha(t-\tau)}C_{vu'}^{\tSS(+)}(-\tau)\nl 
    &\quad -i\sum_{u'}\!\int_0^t\!\!\ud\tau\,c_{\alpha uu'}^{(-)}(t-\tau)e^{i\mu_\alpha(t-\tau)}\Big[ C^{\tSS(-)}_{u'v}(\tau)+C^{\tSS(+)}_{vu'}(-\tau) \Big]\nl 
    &= -i\sum_{u'}\int_0^t\!\!\ud\tau\,g_{\alpha uu'}(t-\tau)e^{i\mu_\alpha(t-\tau)}C^{\tSS(-)}_{u'v}(\tau)\nl 
    &\quad + i\sum_{u'}\int_0^\infty\!\!\ud\tau\,e^{i\mu_\alpha(t+\tau)}\Big[c^{(-)}_{\alpha uu'}(t+\tau)C_{vu'}^{\tSS(+)}(\tau)\nl
    &\qquad\qquad\qquad\qquad\qquad- c^{(+)*}_{\alpha uu'}(t+\tau)C^{\tSS(-)*}_{vu'}(\tau)\Big].
\end{align}
In the same way for $C_{uv}^{\alpha\tS(+)}(t)$, we have 
\begin{align}
    &\quad C_{uv}^{\alpha\tS(+)}(t)\nl 
    &=\la \hat F_{\alpha u}^{(1)\dagger}(t)\hat \psi_v(0)\ra-\la \hat \psi_v(0)\hat F_{\alpha u}^{(2)}(t)\ra +\la\{ \hat F_{\alpha u}^{(2)}(t),\hat \psi_v(0)\}\ra\nl 
    &= i\sum_{u'}\int_0^t\!\!\ud\tau\,g^*_{\alpha uu'}(t-\tau)e^{-i\mu_\alpha(t-\tau)}C^{\tSS(+)}_{u'v}(\tau) \nl 
    &\quad -i\sum_{u'}\int_0^\infty\!\!\ud\tau\,e^{-i\mu_\alpha(t+\tau)}\Big[ c_{\alpha uu'}^{(+)}(t+\tau)C_{vu'}^{\tSS(-)}(\tau)\nl 
    &\qquad\qquad\qquad\qquad\qquad-c_{\alpha uu'}^{(-)*}(t+\tau)C_{vu'}^{\tSS(+)*}(\tau) \Big].
\end{align}
The results can be summarized as 
\begin{align}\label{64}
    &\quad C_{uv}^{\alpha\tS(\sigma)}(t)\nl
    &= \sigma i\sum_{u'}\int_0^t\!\!\ud\tau\,g^{(\sigma)}_{\alpha uu'}(t-\tau)e^{-\sigma i\mu_\alpha(t-\tau)}C^{\tSS(\sigma)}_{u'v}(\tau) \nl 
    &\quad -\sigma i\sum_{u'}\int_0^\infty\!\!\ud\tau\,e^{-i\sigma\mu_\alpha(t+\tau)}\Big[ c_{\alpha uu'}^{(\sigma)}(t+\tau)C_{vu'}^{\tSS(\bar\sigma)}(\tau)\nl 
    &\qquad\qquad\qquad\qquad\qquad-c_{\alpha uu'}^{(\bar\sigma)*}(t+\tau)C_{vu'}^{\tSS(\sigma)*}(\tau) \Big].
\end{align}
Here, we denote $\sigma=+,-$ and $\bar\sigma\equiv-\sigma$  together with $g^{(-)}_{\alpha uu'}(t) \equiv g^{}_{\alpha uu'}(t)$ and $g^{(+)}_{\alpha uu'}(t) \equiv g^{*}_{\alpha uu'}(t)$. Similarly, for the bath--bath correlations, we have 
\begin{align}\label{65}
    &\quad C_{uv}^{\alpha\alpha'(\sigma)}(t) = \delta_{\alpha\alpha'}e^{-\sigma i\mu_\alpha t}c^{(\sigma)}_{\alpha uv}(t) \nl
    & \quad + \sigma i\sum_{u'}\int_0^t\!\!\ud\tau\,g^{(\sigma)}_{\alpha uu'}(t-\tau)e^{-\sigma i\mu_\alpha(t-\tau)}C^{\tS\alpha'(\sigma)}_{u'v}(\tau) \nl 
    &\quad -\sigma i\sum_{u'}\int_0^\infty\!\!\ud\tau\,e^{-i\sigma\mu_\alpha(t+\tau)}\Big[ c_{\alpha uu'}^{(\sigma)}(t+\tau)C_{vu'}^{\alpha'\tS(\bar\sigma)}(\tau)\nl 
    &\qquad\qquad\qquad\qquad\qquad-c_{\alpha uu'}^{(\bar\sigma)*}(t+\tau)C_{vu'}^{\alpha'\tS(\sigma)*}(\tau) \Big].
\end{align}
Equations (\ref{64})  and (\ref{65}) are the SBET for fermionic bath, which relate the entangled system--bath correlations [\Eqs{58a}--(\ref{58d})] and bath--bath correlations [\Eqs{60a} and (\ref{60b})] to  reduced system ones [\Eqs{57a} and (\ref{57b})].  
The former is concerned with the electronic transport current ($t=0$), while the latter is related with the current noise spectrum.

\section{Summary}\label{thsec4}
In summary, this paper extends the system-bath entanglement theorem (SBET), initially established in previous works in terms of linear response functions \cite{Du20034102} and later generalized to correlation functions. \cite{Su23Arxiv2312_13618} 
The focus of this extension relies on the field--dressed scenarios and multiple bosonic Gaussian environments at different temperatures. 
One primary application of the field-dressed SBET lies in the evaluation of nonlinear spectroscopy, for example in pump--probe experiments, using quantum dissipation methods. 
Not only the system but also environments are considered to have optical polarizability, reflecting real-world conditions.
The entangled system--bath correlation functions are crucial in these evaluations.
The established extended SBET
provides an approach to extracting its contribution through the reduced system dynamics.
Thus the paper contributes to the simulation and understanding of entangled system--bath  properties, especially in the context of field--dressed conditions and diverse environmental characteristics, with practical implications in  nonlinear spectroscopies.
Numerical applications will be carried out shortly. 

In addition, we  explore  the relationship between field--free correlation functions and their classical counterparts. 
To provide a comprehensive view, the SBET is also applied to fermionic environments within the transport scenarios.
The extended SBET can be used to reveal the 
intricate interplay between thermal and dynamic processes in the quantum thermoelectric transport.
It is also applicable to thermodynamic and heat transfer evaluations, on such as the hybridization free--energy \cite{Gon20214115,Su23Arxiv2312_13618} and the  heat current.\cite{Du212155,Wan22044102}

\begin{acknowledgements}
	Support from the Ministry of Science and Technology of China (Grant No.\ 2021YFA1200103) and
the National Natural Science Foundation of China (Grant Nos.\ 22103073, 22173088 and 22373091)
is gratefully acknowledged.
\end{acknowledgements}

\appendix

\section{Linear response theory: Quantum versus classical}
\label{thappA}
In this appendix we outline the fundamental quantities and relations of
linear response theory in both quantum and classical statistical mechanics.
To this end, 
we 
pick up the omitted $\hbar$ in our formulas.
In this appendix, we consider the total composite system initially being at the thermal equilibrium,
$\rho^{\rm eq}_{\T}(T)$. Then it is disturbed with $-B\epsilon(t)$.
The measurement object is $\bar A(t)={\rm Tr}[A\rho_{\T}(t)]$.
Here, both $A$ and $B$ are dynamical variables and $\epsilon(t)$ is a real disturbance field.
Interested is
\be
  \delta\bar A(t)=\bar A(t)-\bar A_{\rm eq}={\rm Tr}[\delta A\rho_{\tT}(t)],
\ee
with $\bar A_{\rm eq}={\rm Tr}(A\rho^{\rm eq}_{\tT})$ and $\delta A(t)=A(t)-\bar A_{\rm eq}$.
Both quantum and classical statistical mechanics give the linear response as
\be
  \delta\bar A(t)= \int_{-\infty}^{t}\!\!{\rm d}\tau\,\chi_{AB}(t-\tau)\epsilon(\tau),
\ee
where $\chi_{AB}(t)$ is the response function.

Let us focus on quantum mechanics first, in which $A$ and $B$ are operators.
The response function is resulted as
\be\label{quanresdef}
  \chi_{AB}(t-\tau)\equiv\frac{i}{\hbar}\la[\delta A(t),\delta B(\tau)]\ra
  =\frac{i}{\hbar}\la[A(t-\tau),B(0)]\ra\,.
\ee
Here, $\la \cdots\ra$ is taken over $\rho^{\rm eq}_{\T}(T)$, which is related to the average defined in \Eq{R12} of \Sec{thsec2} in the single--bath and field--free scenarios, with $t_0\rightarrow -\infty$.
Physically, the response function $\chi_{AB}(t)$ is defined for $t\geq 0$.
For $t<0$, it can be formally extended via
\be\label{chitnt}
   \chi_{AB}(-t)=-\chi_{BA}(t).
\ee
The frequency resolution is
\be\label{wchidef}
  \wti\chi_{AB}(\omega)=\int_0^\infty\!{\rm d}t\,e^{i\omega t}\chi_{AB}(t)
  =\wti\chi^{(+)}_{AB}(\omega)+i\wti\chi^{(-)}_{AB}(\omega),
\ee
where
\begin{align}\label{herchiw}
	\begin{split}
        \wti\chi^{(+)}_{AB}(\omega)\equiv\frac{1}{2}\left[\wti\chi_{AB}(\omega)+\wti\chi^\ast_{BA}(\omega)\right]=[\wti\chi^{(+)}_{BA}(\omega)]^\ast,
	\\
        \wti\chi^{(-)}_{AB}(\omega)\equiv\frac{1}{2i}\left[\wti\chi_{AB}(\omega)-\wti\chi^\ast_{BA}(\omega)\right]=[\wti\chi^{(-)}_{BA}(\omega)]^\ast,
	\end{split}
\end{align}
are Hermite and anti-Hermite contributions, respectively.
By \Eq{chitnt}, we have $\wti\chi^{(+)}_{AB}(\omega)=\wti\chi^{(+)}_{BA}(-\omega)$,
$\wti\chi^{(-)}_{AB}(\omega)=-\wti\chi^{(-)}_{BA}(-\omega)$, and
\be
  \wti\chi^{(-)}_{AB}(\omega)=\frac{1}{2i}\int_{-\infty}^\infty\!\!{\rm d}t\,e^{i\omega t}\chi_{AB}(t),
\ee
which serves as the spectral density.

The correlation function is defined as
\be\label{appCdef}
  C_{AB}(t-\tau)\equiv \la\delta A(t)\delta B(\tau)\ra=\la\delta A(t-\tau)\delta B(0)\ra\,.
\ee
It satisfies $C_{AB}(t\rightarrow\infty)=0$,
\be\label{appCABt}
  C^\ast_{AB}(t)=C_{BA}(-t)=C_{AB}(t-i\beta\hbar)\,,
\ee
and
\begin{align}\label{appdotC}
   \dot C_{AB}(t)=\la\delta\dot A(t)\delta B(0)\ra=-\la\delta A(t)\delta\dot B(0)\ra\,.
\end{align}
Similar to the definitions of \Eqs{wchidef} and (\ref{herchiw}),
we can denote $\wti C_{AB}(\omega)$ and $\wti C_{AB}(\omega)=\wti C^{(+)}_{AB}(\omega)+i\wti C^{(-)}_{AB}(\omega)$.
We have
\be\label{appCABw}
  \wti C^{(+)}_{AB}(\omega)=\frac{1}{2}\int_{-\infty}^\infty\!\!{\rm d}t\,e^{i\omega t}C_{AB}(t)=e^{\beta\hbar\w}\wti C^{(+)}_{BA}(-\omega).
\ee
The last identities in \Eq{appCABt} and \Eq{appCABw} accord to the detailed--balance relation.
Related to the response function, obviously, there is $\chi_{AB}(t)=-\frac{2}{\hbar}{\rm Im}C_{AB}(t)$.
We have further
\be\label{appfdt}
    \wti C^{(+)}_{AB}(\omega)=\frac{\hbar\wti\chi^{(-)}_{AB}(\omega)}{1-e^{-\beta\hbar\w}}\,.
\ee
This is the fluctuation--dissipation theorem (FDT) in quantum statistics.
It serves as the fundamental relation between correlation and response functions.

Let us now turn to the classical statistical mechanics in which
$A$ and $B$ are no longer operators and all correlation and response functions are real.
Corresponding relations are in the high-temperature limit of the above quantum ones,
with $\beta\rightarrow0$ [see, e.g., \Eq{appCABt} and \Eq{appCABw}].
The definition of classical correlation function is similar to \Eq{appCdef}
but with the average being over the ensemble distribution.
The classical response function is resulted from classical linear response theory as
\be\label{appclacorr}
  \chi^{\cla}_{AB}(t)=-\beta\dot C^{\cla}_{AB}(t),
\ee
or equivalently,
\be
   \wti\chi^{\cla\,(-)}_{AB}(\omega)=\beta\w\wti C^{\cla\,(+)}_{AB}(\omega),
\ee
which is the hight-temperature limit of \Eq{appfdt}.
The classical real correlation functions satisfy
\be\label{appCABcla}
  C^\cla_{AB}(t)=C^\cla_{BA}(-t),\quad
\big[\wti C^\cla_{AB}(\omega)\big]^\ast=\wti C^\cla_{AB}(-\omega),
\ee
and
\be\begin{split}
&   \big[\wti C^{\cla\,(+)}_{AB}(\omega)\big]^\ast= \wti C^{\cla\,(+)}_{AB}(-\omega)=\wti C^{\cla\,(+)}_{BA}(\omega),\\
&   \big[\wti C^{\cla\,(-)}_{AB}(\omega)\big]^\ast=-\wti C^{\cla\,(-)}_{AB}(-\omega)=\wti C^{\cla\,(-)}_{BA}(\omega).
\end{split}\ee

The correspondence between quantum and classical response functions can also be
easily found via the Kubo transformation,\cite{Kub57570} by which, the quantum response function [\Eq{quanresdef}]
can be rewritten as
\be\label{appkubo}
\chi_{AB}(t) = -\beta \la \dot{\ti A}(t)B(0)\ra,
\ee
with
\be\label{apptiA}
\ti A(t) = \frac{1}{\beta}\int_{0}^{\beta}\!{\rm d}\lambda\,e^{\lambda H_{\T}}A(t) e^{-\lambda H_{\\T}}.
\ee
This transformation can be verified via the Kubo identity
\be
[\hat A(t), e^{-\beta H_{\T}}]=-e^{-\beta H_{\T}}\!\!\int_{0}^{\beta}\!{\rm d}\lambda\,e^{\lambda H_{\T}}[\hat A(t),H_{\T}] e^{-\lambda H_{\T}}.
\ee
In the classical limit, operators correspond to physical observables commuting with each other,
leading to \Eq{apptiA}, $\ti A^{\cla}(t)=A^{\cla}(t)$.
Thus, \Eq{appkubo} reduces to the classical correspondence, \Eq{appclacorr}.
The classical response functions can also be written as [cf.\ \Eq{appdotC}]
\be\label{appfinalt}
      \chi^\cla_{AB}(t) = -\beta      C^\cla_{\dot A B}(t) =\beta      C^\cla_{A \dot B}(t)   .
\ee

\section{Classical limit of bosonic field--free SBET}\label{thappB}

This appendix presents the classical limit of the bosonic field--free SBET. Firstly, we give the field--free version of the extended SBET. We recast \Eq{eq17} as $R_{O_2O_1}(t_2,t_1)=iC_{O_2O_1}(t_2-t_1)$ where $C_{O_2O_1}(t)$ is defined with respect to the steady state in the limit of $t_0\to-\infty$. We obtain from the extended SBET, 
\begin{align}\label{gSBET1}
    &C_{F_{\alpha v}Q_{v'}}(t) = \sum_{v''}\int_{0}^{t}\!\d\tau\,\phi_{\alpha vv''}(t-\tau)\eta_{\alpha v''}C_{Q_{v''}Q_{v'}}(\tau) \nl 
    &\quad- \frac{2}{\hbar}{\rm Im} \Bigg\{\sum_{v''}\int^{\infty}_{0}\!\ud\tau\,c_{\alpha vv''}(t+\tau)\eta_{\alpha v''}C_{Q_{v'}Q_{v''}}(\tau)\Bigg\}\nl 
    &=[C_{Q_{v'}F_{\alpha v}}(-t)]^*,
\end{align}
and 
\begin{align}\label{gSBET2}
    &C_{F_{\alpha v}F_{\alpha'v'}}(t) = \delta_{\alpha\alpha'}c_{\alpha vv'}(t)\nl 
    &\ + \sum_{v''}\int_{0}^{t}\!\d\tau\,\phi_{\alpha vv''}(t-\tau)\eta_{\alpha v''}C_{Q_{v''}F_{\alpha'v'}}(\tau) \nl 
    &\ - \frac{2}{\hbar}{\rm Im} \Bigg\{\sum_{v''}\int^{\infty}_{0}\!\ud\tau\,c_{\alpha vv''}(t+\tau)\eta_{\alpha v''}C_{F_{\alpha'v'}Q_{v''}}(\tau)\Bigg\}.
\end{align}
These are the multi--bath extensions of the SBET in Ref.\,\onlinecite{Su23Arxiv2312_13618}.

The classical limit is obtained by using the Wigner representation and expanding the functions up to the first order of $\hbar$. 
In the Wigner representation, the basic algebra of operators reads 
\begin{align}\label{ABW}
    \big(\hat A\hat B\big)_{W}(p,q) = A_W(p,q) \exp\bigg( \frac{i\hbar}{2}\TT \bigg)B_W(p,q),
\end{align}
where 
\begin{align}
    \TT \equiv \sum_{k}\frac{\Lpartial}{\partial q_k}\frac{\Rpartial}{\partial p_k} - \frac{\Lpartial}{\partial p_k}\frac{\Rpartial}{\partial q_k}.
\end{align}
We have for the correlation function, 
\begin{align}\label{Ccl}
    &\quad C_{AB}(t)\nl 
    &= \big\la A_W(t)B_W(0)\big\ra_{\rm cl} + \frac{i\hbar}{2}\big\la\{A_W(t),B_W(0)\}_{\textsc{pb}}\big\ra_{\rm cl} + \mathcal{O}(\hbar^2)\nl 
    &\equiv C_{AB}^{\rm cl}(t) - \frac{i\hbar}{2}\chi_{AB}^{\rm cl}(t)+ \mathcal{O}(\hbar^2),
\end{align}
where the Poisson bracket is
\begin{align}\label{appbpoisson}
    \{A_W,B_W\}_{\PB} \equiv \sum_k \bigg(\frac{\partial A_W}{\partial q_k}\frac{\partial B_W}{\partial p_k} - \frac{\partial A_W}{\partial p_k}\frac{\partial B_W}{\partial q_k}\bigg)
\end{align}
and the average is 
\begin{align}
    \la A_W(t)B_W(0)\ra_{\rm cl} \equiv \int\!A_W(t)B_W(0)\rho_W^{\rm st}\prod_k\frac{\ud p_k\ud q_k}{2\pi\hbar}
\end{align}
with $\rho_W^{\rm st}$ being the steady state in the Wigner representation. In term of the Poisson bracket [\Eq{appbpoisson}], the dynamic observable satisfies 
\begin{align}
    \dot{A}_W(t) = \{A_W(t), H_W\}_{\PB} + \mathcal{O}(\hbar^2).
\end{align}
In \Eq{Ccl}, 
\begin{align}
    \chi_{AB}^{\rm cl}(t) \equiv -\big\la\{A_W(t),B_W(0)\}_{\textsc{pb}}\big\ra_{\rm cl}.
\end{align}
From \Eqs{quanresdef} and (\ref{ABW}), we have 
\begin{align}\label{chicl}
    \chi_{AB}(t)
    &= \chi_{AB}^{\rm cl}(t) + \mathcal O(\hbar^2).
\end{align}
For later use, we denote 
\begin{subequations}\label{B11}
    \begin{align}
    C_{Q_vQ_{v'}}(t) &\equiv C^{\tSS;\rm cl}_{vv'}(t) - \frac{i\hbar}{2}\chi^{\tSS;\rm cl}_{vv'}(t) + \mathcal{O}(\hbar^2),\\
    C_{F_{\alpha v}Q_{v'}}(t) &\equiv C^{\alpha\tS;\rm cl}_{vv'}(t) - \frac{i\hbar}{2}\chi^{\alpha\tS;\rm cl}_{vv'}(t) + \mathcal{O}(\hbar^2),\\
    C_{F_{\alpha v}F_{\alpha'v'}}(t) &\equiv C^{\alpha\alpha';\rm cl}_{vv'}(t) - \frac{i\hbar}{2}\chi^{\alpha\alpha';\rm cl}_{vv'}(t) + \mathcal{O}(\hbar^2),
    \end{align}
\end{subequations}
and 
\begin{align}\label{B12}
    c_{\alpha vv'}(t) = c_{\alpha vv'}^{\rm cl}(t) - \frac{i\hbar}{2}\phi_{\alpha vv'}^{\rm cl}(t) + \mathcal{O}(\hbar^2).
\end{align}
In \Eqs{gSBET1} and (\ref{gSBET2}), 
\begin{align}\label{B13}
    \phi_{\alpha vv'}(t) = \phi_{\alpha vv'}^{\rm cl}(t) + \mathcal O(\hbar^2)
\end{align}
is similar as \Eq{chicl}.

Substituting \Eqs{B11}-(\ref{B13}) into \Eqs{gSBET1} and (\ref{gSBET2}) followed by $\hbar\to 0$, we obtain 
\begin{align}\label{gSBET1cl}
    &C_{vv'}^{\alpha\tS;\rm cl}(t) = \sum_{v''}\int_{0}^{t}\!\d\tau\,\phi_{\alpha vv''}^{\rm cl}(t-\tau)\eta_{\alpha v''}C_{{v''}{v'}}^{\tSS;\rm cl}(\tau) \nl 
    &\qquad\qquad\quad+\!\sum_{v''}\!\int_0^\infty\!\!\!\ud\tau\,\eta_{\alpha v''}\Big[ c^{\rm cl}_{\alpha vv''}(t+\tau)\chi^{\tSS;\rm cl}_{v'v''}(\tau)\nl 
    &\qquad\qquad\qquad+\phi^{\rm cl}_{\alpha vv''}(t+\tau)C^{\tSS;\rm cl}_{v'v''}(\tau) \Big]
\end{align}
and 
\begin{align}\label{gSBET2cl}
    C^{\alpha\alpha';\rm cl}_{vv'}(t) &= \delta_{\alpha\alpha'}c^{\rm cl}_{\alpha vv'}(t)\nl 
    &\ \ + \sum_{v''}\int_{0}^{t}\!\d\tau\,\phi^{\rm cl}_{\alpha vv''}(t-\tau)\eta_{\alpha v''}C^{\tS\alpha';\rm cl}_{{v''}{v'}}(\tau) \nl 
    &\ \ +\sum_{v''}\int_0^\infty\!\!\ud\tau\,\eta_{\alpha v''}\Big[ \phi^{\rm cl}_{\alpha vv''}(t+\tau)C^{\alpha'\tS;\rm cl}_{v'v''}(\tau)\nl 
    &\qquad +c^{\rm cl}_{\alpha vv''}(t+\tau)\chi^{\alpha'\tS;\rm cl}_{v'v''}(\tau) \Big].
\end{align}
These are the classical SBET
for any steady states. They intimately relate to the molecular dynamics simulation. For the single bath at the thermal equilibrium with \Eq{appclacorr} existing, by taking the time derivative on both sides of \Eqs{gSBET1cl} and (\ref{gSBET2cl}), the contributions from the $\int_0^\infty$--integral--terms vanish and \Eqs{gSBET1cl} and (\ref{gSBET2cl}) lead to the relations between response functions being of the same forms as the quantum SBET in Ref.\,\onlinecite{Du20034102}. So far, we have verified it via molecular dynamics for the Brownian oscillator system.

\section{Fermionic Bogoliubov transformation
}\label{thappC}

Bogoliubov transformation for free fermions is introduced in detail here.
%
By the Bogoliubov transformation in thermofield method, we could construct the effective zero temperature bath.
Firstly, we add an assistant bath  ($\{\hat a_{\alpha j}^{\prime\dg},\hat a'_{\alpha j}\}$) for each reservoir, which does not affect the original system--and--environment dynamics. It results in
\begin{align} \label{eb}
  \ti h^{\rm eff}_{\alpha}&=\sum_{j}(\epsilon_{\alpha j}-\mu_\alpha)\big(\hat a_{\alpha j}^{\dg}\hat a_{\alpha j}-\hat a_{\alpha j}^{\prime\dg}\hat a'_{\alpha j}\big).
\end{align}
It can be recast as
\begin{align} \label{eb2}
  \ti h^{\rm eff}_{\alpha}
  =\sum_{j}(\epsilon_{\alpha j}-\mu_\alpha)\big(\hat d_{\alpha j}^{\prime\dg}\hat d^{\prime}_{\alpha j}-\hat d_{\alpha j}^{\dg}\hat d_{\alpha j}\big),
\end{align}
with  the transformation
\be \label{bt}
\begin{split}
\hat  a_{\alpha j}=\sqrt{1-\bar{n}_{\alpha j}}\hat d'_{\alpha j}+\sqrt{\bar{n}_{\alpha j}}\hat d_{\alpha j}^{\dagger}, 
\\  
\hat a'_{\alpha j}=\sqrt{1-\bar{n}_{\alpha j}}\hat d_{\alpha k}- \sqrt{\bar{n}_{\alpha j}}\hat d_{\alpha j}^{\prime\dagger},
\end{split}
\ee
and its inverse
\be
\begin{split}\label{ibt}
\hat d_{\alpha j}=\sqrt{1-\bar{n}_{\alpha j}}\hat a'_{\alpha j}+ \sqrt{\bar{n}_{\alpha j}}\hat a_{\alpha j}^{\dagger},
\\
\hat d'_{\alpha j}=\sqrt{1-\bar{n}_{\alpha j}}\hat a_{\alpha j}-\sqrt{\bar{n}_{\alpha j}}\hat a_{\alpha j}^{\prime\dagger}.
\end{split}
\ee
Here,  $\bar{n}_{\alpha j}=[e^{\beta_{\alpha}(\epsilon_{\alpha j}-\mu_{\alpha})}+ 1]^{-1}$ is the average occupation number for the Fermi--Dirac distribution.
Equation (\ref{eb2}) can be easily verified via  \Eqs{bt}  and (\ref{ibt}), with noticing
$
    \{\hat  a_{\alpha j},\hat a_{\alpha' j'}^{\dagger}\}=\{\hat  a'_{\alpha j},\hat a_{\alpha' j'}^{\prime\dagger}\}=\{\hat  d_{\alpha j},\hat d_{\alpha' j'}^{\dagger}\}=\{\hat  d'_{\alpha j},\hat d_{\alpha' j'}^{\prime\dagger}\}=\delta_{\alpha\alpha'}\delta_{jj'}
$,\, $
    \{\hat  a_{\alpha j},\hat a_{\alpha' j'}^{}\}=\{\hat  a'_{\alpha j},\hat a_{\alpha' j'}^{\prime}\}=\{\hat  d_{\alpha j},\hat d_{\alpha' j'}^{}\}=\{\hat  d'_{\alpha j},\hat d_{\alpha' j'}^{\prime}\}=0
$, and $
    \{\hat  a_{\alpha j},\hat a'_{\alpha' j'}\}=\{\hat  a_{\alpha j},\hat a_{\alpha' j'}^{\prime\dagger}\}=\{\hat  d_{\alpha j},\hat d'_{\alpha' j'}\}=\{\hat  d_{\alpha j},\hat d_{\alpha' j'}^{\prime\dagger}\}=0
$.
Accordingly, the system--bath interacting term can be recast into 
\begin{align}
 \!\! \sum_{\alpha u}\big( \hat \psi_u^\dagger\hat F_{\alpha u}+{\rm h.c.} \big) = \sum_{\alpha u}\big[\hat \psi_{u}^{\dg}\big(\hat F^{(1)}_{\alpha u}+\hat F^{(2)\dg}_{\alpha u}\big)+{\rm h.c.}\big]
\end{align}
with
\bsube
\be
\hat F^{(1)}_{\alpha u}=\sum_j t^{*}_{\alpha u j}\sqrt{1-\bar{n}_{\alpha j}}\hat d'_{\alpha j},
\ee
and
\be
\hat F^{(2)}_{\alpha u}=\sum_j t_{\alpha u j}\sqrt{\bar{n}_{\alpha j}}\hat d_{\alpha j}.
\ee
\esube

As known, a mixed state may be viewed as the reduction of one pure state in a higher dimensional Hilbert space. Such a pure state is called the purification of the original mixed state. We now try to find the purification of the thermal state. 
For simplicity, we go with the single--mode  ($\epsilon_{\alpha j}$) case. The thermal state reads
\begin{align}\label{rhok}
    \rho_{\alpha j}(\beta_{\alpha},\mu_{\alpha})\!&=\frac{e^{-\beta_{\alpha} (\epsilon_{\alpha j}-\mu_{\alpha}) \hat a_{\alpha j}^{\dagger}\hat a_{\alpha j}}}{Z_{\alpha j}}
\nl &  
    =\sum_{n_{\alpha j}=0,1} \!\!\!\!\frac{e^{-\beta_{\alpha} n_{\alpha j}(\epsilon_{\alpha j}-\mu_\alpha)}}{Z_{\alpha j}}|n_{\alpha j}\rangle \langle n_{\alpha j}|, 
\end{align}
where the partition function $Z_{\alpha j}=1+e^{-\beta_{\alpha}(\epsilon_{\alpha j}-\mu_\alpha)}$.
The purification of the mixed state in \Eq{rhok} reads
\begin{align}
    |\xi_{\alpha j} \rangle=\frac{1}{\sqrt{Z_{\alpha j}}}\sum_{n_{\alpha j}=0,1}e^{-\beta_{\alpha} n_{\alpha j}(\epsilon_{\alpha j}-\mu_\alpha)/2}|n_{\alpha j}\rangle |n_{\alpha j}\rangle'.
\end{align}
Here, we denote $|n_{\alpha j}\rangle \equiv (\hat a_{\alpha j}^{\dagger})^{n_{\alpha j}}|0_{\alpha j}\rangle $ and $|n_{\alpha j}\rangle' \equiv (\hat a_{\alpha j}^{\prime\dagger})^{n_{\alpha j}}|0_{\alpha j}\rangle'$.
It is easy to verify that 
\begin{align}
    \rho_{\alpha j}={\rm tr}'\big(|\xi_{\alpha j} \ra\la\xi_{\alpha j}|\big),
\end{align}
where ${\rm tr}'$ represents the partial trace with respect to the assistant bath degrees of freedom.
For the multi--mode cases, the density operator and the corresponding purification are just the direct product of the single--mode ones. That is to say, 
$|\xi_{\alpha}\rangle=\prod_{j}\otimes|\xi_{\alpha j} \rangle$
and
$\rho_{\alpha}=\prod_{j}\otimes\rho_{\alpha j}$.
This finishes the purification of the grand canonical thermal state of each reservoir.
It is easy to find that
\begin{align}\label{tbf1}
    \hat d_{\alpha j}|\xi_{\alpha j} \rangle=\hat d'_{\alpha j}|\xi_{\alpha j}\ra =0.
\end{align}
Thus 
\be |\xi\ra\la\xi|\equiv\prod_{\alpha j}|\xi_{\alpha j}\ra\la\xi_{aj}|
\ee
is indeed the vacuum state of $\ti h^{\rm eff}_{\alpha}$.


\end{document}